\documentclass[11pt,fleqn]{article}

\usepackage{a4wide}
\usepackage{cite}
\usepackage{epsf}
\usepackage{rotate}

\renewcommand{\thefootnote}{\fnsymbol{footnote}}    

\begin{document}

\thispagestyle{empty}

\begin{flushright}
TUM-T31-98/96 \\
hep-ph/9611406 \\
September 1996 \\
\end{flushright}

\begin{center}

\vskip 2cm
\Large
{\bf $1/m_b^2$ correction to the left-right lepton polarization
asymmetry in the decay $B \rightarrow X_s \, \mu^+ \mu^-$}
\vskip 1.5cm

\large
Oliver B{\"a}r and Nicolas Pott\footnote[2]{
Supported by the German Bundesministerium f{\"u}r Bildung und Forschung
under contract 06 TM 743 and DFG Project Li 519/2-1.}

\vskip 1cm
\small
{\em Physik-Department, Technische Universit{\"a}t M{\"u}nchen \\
 D-85748 Garching, Germany}
\end{center}

\vskip 2cm
\begin{abstract}
Using a known result by Falk {\em et al.} for the $1/m_b^2$ correction to
the dilepton invariant mass spectrum in the decay $B \rightarrow X_s
\, \mu^+ \mu^-$, we calculate the $1/m_b^2$
correction to the left-right muon polarization asymmetry in this
decay.
Employing an up-to-date range of values for
the non-perturbative parameter $\lambda_1$, we find that the
correction is much smaller than it should have been expected from
the previous work by Falk {\em et al.}
\end{abstract}

\vspace{3cm}
\begin{center}
To appear in {\em Physical Review D}.
\end{center}

\renewcommand{\thefootnote}{\arabic{footnote})}

\newpage
\pagenumbering{arabic}

\subsubsection*{Introduction}

Rare decays of $B$ mesons have been studied extensively in the last
few years. Such decays are forbidden in the tree-level
approximation and proceed through loop diagramms only. Consequently,
they are sensitive to the complete particle content of a given theory
and their analysis may thus shed some light on possible new physics
beyond the standard model (SM).  In this paper, we focus on the
inclusive decay $B
\rightarrow X_s \, \mu^+ \mu^-$, where $X_s$ denotes an arbitrary
hadronic final state with total strangeness $-1$\footnote{Since we
neglect the mass of the leptons, all statements in this paper apply
likewise to the decay $B \rightarrow X_s \, e^+e^-$. However, this
decay mode is experimentally even harder accessible than $B
\rightarrow X_s \, \mu^+\mu^-$, so we decided to mention only the latter
one explicitly in the text.}. In contrast to the decay $B\rightarrow X_s
\,\gamma$, the three-body decay $B\rightarrow X_s \, \mu^+\mu^-$
allows one to define and measure several kinematic
distributions. Beside the invariant mass spectrum of the lepton pair
 a
forward-backward charge asymmetry \cite{alietalfb} and a left-right
polarization asymmetry \cite{hewett,liudelb} have been proposed. The simultaneous
measurement of these distributions allows one to extract the values of
the Wilson coeffients which govern the decay amplitude and contain the
short distance physics \cite{alietalc67}. The values of these Wilson coefficients are
sensitive to new physics and in a measurement of the kinematic
distributions one might be able to detect deviations from the SM
prediction. For this reason the distributions
mentioned above should be calculated as precise as possible.

Meanwhile a complete next-to-leading order (NLO) calculation of the
relevant Wilson coefficients is available
\cite{misiakb393,burasmuenz}. These coefficients properly include the short
distance QCD effects and the NLO approximation considerably reduces
the (otherwise significant) theoretical uncertainty in the final
result which stems from its dependence on the renormalization scale.

Apart from these perturbative calculations which rely entirely on the
spectator model (defined as the free quark decay model including
perturbative QCD),  long distance
effects due to the non-perturbative physics at low scales have to be
taken into account. Important corrections are first of all generated
by intermediate $c\bar{c}$ bound states (e.\,g. the channel $b
\rightarrow s \, J/\psi \rightarrow s \mu^+\mu^-$) which
influence the kinematic distributions even far away from the resonance
region. For a way to deal with these effects we refer to the
literature \cite{lim,deshpandepanose,donnel,kruegersehgal}.

Additionally, there are non-perturbative effects due to the binding of
the $b$ quark inside the $B$ meson. It is strongly expected that for inclusive
decays of heavy quarks such effects may be incorporated by a
controlled expansion in inverse powers of the mass of the decaying
quark, the so-called heavy quark expansion (HQE)
\cite{chayetal,bigietal}\footnote{Whether this expansion can 
be justified theoretically rigorous is still the subject of ongoing
discussions \cite{chibisov}. At the moment,
phenomenology seems to be the only way to judge the validity
of the HQE.}. The 
first term of this expansion can be shown to reproduce the spectator
model. Falk, Luke and Savage \cite{falklukesavage94} calculated the first
non-vanishing -- i.\,e. $O(1/m_b^2)$ -- correction to the dilepton
invariant mass spectrum in the decay $B \rightarrow X_s \, \mu^+
\mu^-$ some time ago.

It is this type of corrections we are dealing with in the following.
We slightly extend the work of Liu and Delbourgo \cite{liudelb} about the
left-right polarization asymmetry by presenting the
$1/m_b^2$ correction to this distribution. It should particularly be stressed
that in the numercial analysis we take into account a present range of
values for $\lambda_1$ (one of the non-perturbative parameters
that occur by performing the HQE), and that these values differ
drastically from the one used in Ref.\ \cite{falklukesavage94}.

\subsubsection*{The left-right asymmetry}

The general framework for decays like $B\rightarrow X_s\,\mu^+\mu^-$ is
the effective theory approach 
with the effective Hamiltonian as the central element. Once the
effective Hamiltonian is given one can calculate the decay amplitude
and the kinematic distributions mentioned above. The effective
Hamiltonian for the decay $B\rightarrow X_s\,\mu^+\mu^-$ has been
calculated in the NLO approximation by Misiak \cite{misiakb393}
and independently by Buras and M{\"u}nz \cite{burasmuenz}. In our paper we entirely use the
notation introduced in Ref.\ \cite{burasmuenz} and we refer to this paper for the operator
basis, the analytic expressions of the Wilson coefficients and for the
details concerning the NLO calculation.

Let us now consider the left-right asymmetry. It is defined as
\begin{eqnarray}
\frac{dA^{LR}}{d\hat{s}} & = &  \frac{d\Gamma^L}{d\hat{s}} -  
\frac{d\Gamma^R}{d\hat{s}},\nonumber
\end{eqnarray}
where
\begin{eqnarray}
\hat{s} & = & \frac{(p_{\mu^+} + p_{\mu^-})^2}{m_b^2}\nonumber
\end{eqnarray} 
is the scaled 4-momentum transfer to the muons and
$d\Gamma^L/d\hat{s}$\, ($d\Gamma^R/d\hat{s}$) denotes the
invariant mass spectrum for a decay into purely left-handed (right-handed) muons.
Instead of calculating $d\Gamma^L/d\hat{s}$\ and $d\Gamma^R/d\hat{s}$
separately, $dA^{LR}/d\hat{s}$ can alternatively be obtained directly from the
invariant mass spectrum $d\Gamma/d\hat{s}$. With the operator basis used
in Ref. \cite{burasmuenz}, $d\Gamma/d\hat{s}$ is derived in terms of the Wilson
coefficients  $C_9$ and $C_{10}$. The corresponding operator $Q_9$
contains a vector coupling of the muons, $Q_{10}$ an axial-vector
coupling. Defining a left-handed and a right-handed operator by
\begin{eqnarray}
Q_L & = & Q_9 - Q_{10} , \nonumber\\
Q_R & = & Q_9 + Q_{10},\nonumber
\end{eqnarray}
one obtains $d\Gamma/d\hat{s}$ in terms of the corresponding Wilson
coefficients $C_L$ and $C_R$ by the
simple substitution
\begin{eqnarray}
C_9 & = & C_L+C_R,\nonumber\\
C_{10} & = & C_R-C_L.\nonumber
\end{eqnarray}
Since the remaining operators $Q_1 \cdots Q_8$ contain only vector
couplings to the muons, one finds in this basis
\begin{eqnarray}
\frac{dA^{LR}}{d\hat{s}} & = & \left.
\frac{d\Gamma}{d\hat{s}}\right|_{C_R=0} -\left.
\frac{d\Gamma}{d\hat{s}}\right|_{C_L=0} .   
\nonumber
\end{eqnarray}
The final result may be transformed back to the basis $Q_9$,
$Q_{10}$ and re-expressed in terms of $C_9$ and $C_{10}$\footnote
{Note that 
\begin{eqnarray}
\left. \frac{d\Gamma}{d\hat{s}}\right|_{C_R=0}\neq\frac{d\Gamma^L}{d\hat{s}}&
\mbox{and} &\left. \frac{d\Gamma}{d\hat{s}}\right|_{C_L=0}\neq\frac{d\Gamma^R}{d\hat{s}}\nonumber
\end{eqnarray}
because $\left. \frac{d\Gamma}{d\hat{s}}\right|_{C_R=0}$ contains
right-handed components due to the operators $Q_1\cdots Q_8$ which
cancel against the same terms present in $\left.
\frac{d\Gamma}{d\hat{s}}\right|_{C_L=0}$ by taking the difference.}.
Following these steps  one reproduces the result for the left-right
asymmetry in Ref.\ \cite{liudelb} if one starts with the invariant
mass spectrum as given in 
Ref.\ \cite{burasmuenz}.
Keeping this procedure in mind one can extract the $1/m_b^2$-correction
to the left-right asymmetry from the 
$1/m_b^2$-correction to the invariant mass spectrum given in
\cite{falklukesavage94}.
We stress that it is not necessary to repeat the full calculation
Falk {\em et al.} have done in order to find the $1/m_b^2$-correction for
the left-right asymmetry\footnote{This is not the case for the
$1/m_b^2$-correction to the forward-backward asymmetry, which cannot
be extracted from the result in \cite{falklukesavage94}.}.
We obtain
\begin{eqnarray}
\frac{dA^{LR}}{d\hat{s}} & = & \frac{dA_0^{LR}}{d\hat{s}}\, + \, \frac{dA_2^{LR}}{d\hat{s}},\nonumber\\
\frac{dA_0^{LR}}{d\hat{s}} & = & -\,\frac{\Gamma(b \rightarrow c l^-
\overline{\nu})}{f(z)
\cdot\kappa(z)}|V_{tb}|^2\,\left|\frac{V_{ts}}{V_{cb}}\right|^2
\,\frac{\alpha^2}{2\pi^2} \, (1-\hat{s})^2 \tilde{C}_{10}\left[(1+2\hat{s})\mbox{Re}\left(\tilde{C}_9^{eff} \right)\, +\,
6\,C_7^{(0)eff}\right],
\nonumber\\
\frac{dA_2^{LR}}{d\hat{s}} & = & -\,\frac{\Gamma(b \rightarrow c l^-
\overline{\nu})}{f(z)
\cdot\kappa(z)}|V_{tb}|^2\,\left|\frac{V_{ts}}{V_{cb}}\right|^2
\,\frac{\alpha^2}{2\pi^2} \, (1-\hat{s})\tilde{C}_{10}
\left(k_1^{LR}(\lambda_1,\hat{s}) +
k_2^{LR}(\lambda_2,\hat{s})\right),\nonumber
\end{eqnarray}
where we defined $z = m_c/m_b$ and 
\begin{eqnarray}
k_1^{LR}(\lambda_1,\hat{s}) & = &  \frac{\lambda_1}{2m_b^2} \left\{ \left(-2\hat{s}^2+3\hat{s}+5\right)\mbox{Re}\left(\tilde{C}_9^{eff}\right) -2(3\hat{s} -5)C_7^{(0)eff}\right\},\nonumber\\
k_2^{LR}(\lambda_2,\hat{s}) & = &   \frac{\lambda_2}{2m_b^2}\left\{3\left(-10\hat{s}^2 +15 \hat{s} +1 \right)\mbox{Re}\left(\tilde{C}_9^{eff}\right)-6(7\hat{s} -5)C_7^{(0)eff} \right\}.\nonumber
\end{eqnarray}
$dA_0^{LR}/d\hat{s}$ denotes the spectator model result. It agrees with the
result in Ref.\ \cite{liudelb}. The functions $f(z)$ and $\kappa(z)$ are the
phase-space factor and the one gluon correction to the decay
$b\rightarrow c e \bar{\nu}$ given in Eqn. (2.31)
and (2.32) of Ref.\ \cite{burasmuenz}. The Wilson coefficients
$C_7^{(0)eff},\,\tilde{C}_9^{eff}\mbox{and } \tilde{C}_{10}$ are also
defined in Ref.\ \cite{burasmuenz}, Eqn. (2.3), (2.28) and (2.8). Note
that the effective coefficient  $\tilde{C}_9^{eff}$ depends
explicitely on $\hat{s}$ because it contains the influence of the operators
$Q_1 \cdots Q_6$.  
Finally, $dA_2^{LR}/d\hat{s}$ denotes the
$1/m_b^2$-correction, given in terms of the two HQET parameter
$\lambda_1$ and $\lambda_2$. A definition of them as matrix elements
can be found in Ref.\ \cite{falklukesavage94}, Eq. (2.21).  

\subsubsection*{Numerics}

Let us now examine the size of the
correction compared with the spectator 
model result. In doing this we do not incorporate the long distance
effects due to intermediate $c\bar{c}$ resonances. We do not neglect them because
they are small -- in fact, as already mentioned, they do have a large
effect on the shape of the spectra even far away from the resonance
region. The common way to deal with these resonances is a simple
replacement of  $\tilde{C}_9^{eff}$ by $\tilde{C}_9^{eff} +
\mbox{Res($\hat{s}$)}$ (see e.\,g. Ref.\ \cite{kruegersehgal}), where
the function Res$(\hat{s})$ is chosen to  produce resonance peaks at
the masses of 
the intermediate $c\bar{c}$ bound states.
The relative correction to the
spectator model, however,  is nearly the same with or without this
replacement of $\tilde{C}_9^{eff}$. So,
for simplicity, we show the non-resonant results only. 

In the appendix we give a list of the numerical values of all input
parameter we used.  Some remarks should be made concerning the
non-perturbative parameter $\lambda_1$ and $\lambda_2$. From the
$B\,-\,B^*$ mass splitting the numerical value of $\lambda_2$ is well
known to be 0.12$\,\mbox{GeV}^2$ \cite{neubertrev}. $\lambda_1$ is
much less understood. In Ref.\ \cite{falklukesavage94} $\lambda_1 =
0.5\,\mbox{GeV}^2$ was used in the numerical analysis. These values
led to an enhancement of roughly 10 percent over the full range of
$\hat{s}$ in the invariant mass spectrum. This value of $\lambda_1$,
however, is no longer appropriate. Even if there is still some
controversary about the magnitude, the sign of $\lambda_1$ is strongly
believed to be negative.  There have been attempts to calculate
$\lambda_1$ using QCD sum rules
\cite{neubert1992,eletskii,ballbraun,neubert9608211}
and, alternatively, to extract it by a fit to the
data from $B$ and $D$ decays \cite{falketal95,gremmetal}.  Even an
upper bound has been proposed \cite{bigietal94,bigietal95,kapustin}.
However, the values found lie in the range
$-0.6\,\mbox{GeV}^2\leq\,\lambda_1\,\leq\,-0.1\,\mbox{GeV}^2$, which
is quite large.

In Fig.\,1 we plot the spectator model prediction with and without the
$1/m_b^2$ correction. 
\begin{figure}[thb]
\centerline{
\epsfysize=11cm
\rotate[r]{
\epsffile{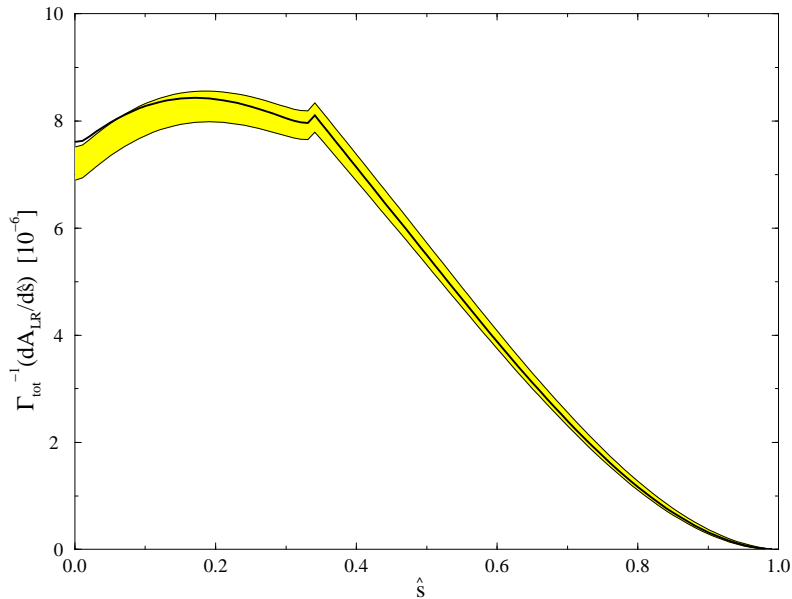}}
}
\caption{\small The left-right asymmetry in the decay $B \rightarrow
X_s \mu^+ \mu^-$ (normalized to the total decay
width of the $B$ meson). The solid line represents the spectator model
result. The shaded area corresponds to the left-right asymmetry
including the $1/m_b^2$ correction, varying $\lambda_1$ between
$-0.6\,\mbox{GeV}^2$ and $-0.1\,\mbox{GeV}^2$.
The upper edge of the area is given by $\lambda_1
\,=\,-0.1\,\mbox{GeV}^2$, the lower edge by $\lambda_1
\,=\,-0.6\,\mbox{GeV}^2$.}  
\label{fig1}
\end{figure}
The influence of the correction never exceeds a rate of 3.5\%
in the high $\hat{s}$-region $(\hat{s}\geq 0.4)$
and maximally reduces the spectrum by 9.5\% for very small values of $\hat{s}$.
We emphasize that the smallness of the correction is due to an
accidental cancellation of the corrections $k_1^{LR}$ and $k_2^{LR}$. With
$\lambda_1 \approx -3\,\lambda_2$ the functions $k_1^{LR}$ and
$k_2^{LR}$ almost cancel in the high $\hat{s}$ region.
For the integrated left-right asymmetry the correction lies between
+3\% and --5\% compared to the spectator model result. 

This small correction in mind it is interesting to show again the
$1/m_b^2$ correction to the invariant mass distribution with 
negative values of $\lambda_1$. This is done in Fig.\,2.
\begin{figure}[tbh]
\centerline{
\epsfysize=11cm
\rotate[r]{
\epsffile{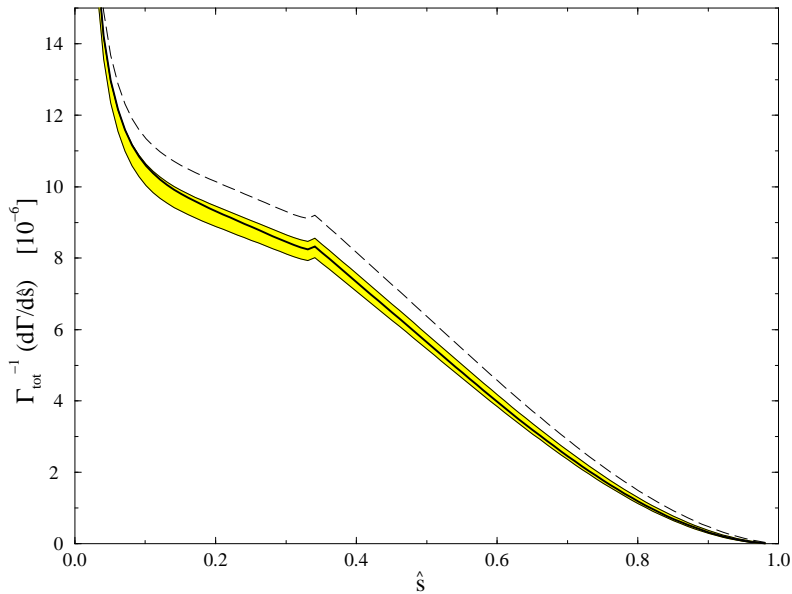}}
}
\caption{\small The dilepton invariant mass spectrum  in the decay $B
\rightarrow 
X_s \mu^+ \mu^-$ (normalized to the total decay
width of the $B$ meson). The solid line represents the spectator model
result. The shaded area corresponds to the spectrum
including the $1/m_b^2$ correction, varying $\lambda_1$ between
$-0.6\,\mbox{GeV}^2$ and $-0.1\,\mbox{GeV}^2$. As in Fig. 1 the upper
edge of the area is given by $\lambda_1\,=\,-0.1\,\mbox{GeV}^2$, the
lower edge by $\lambda_1\,=\,-0.6\,\mbox{GeV}^2$.  The
dashed line is the result obtained by Falk {\em et al.} using
$\lambda_1=0.5\,\mbox{GeV}^2$.}
\label{fig2}
\end{figure}
The overall correction is significantly reduced compared to the result in
Ref.\ \cite{falklukesavage94}. As in the case for the left-right asymmetry, the
correction to the spectator model is at most 3.5\% for
$\hat{s}\geq0.4$ and
maximally differs by 5.2\% near $\hat{s} = 0.1$.
The correction results in a maximal decrease of the
integrated spectrum of 3.8\% 
for $\lambda_1 =-0.6\,\mbox{GeV}^2$. 
In comparison, the integrated spectrum increases by 9.3\% taking
$\lambda_1$  as high as the value used in Ref.\ \cite{falklukesavage94}.

As long as $\lambda_1$ is poorly known, the following numerical formulae might
be useful. They  describe the relative
size 
of the $1/m_b^2$ correction to the spectator model results of the
integrated left-right asymmetry and the branching ratio:
\begin{eqnarray}
\frac{\Delta A^{LR}}{A^{LR}_0} & = &
\left(\left[\frac{\lambda_1}{\mbox{GeV}^2}\right]\, + \,0.28 \right) \times
15.3\,\%\, ,\nonumber\\
\frac{\Delta {\cal B}[B\rightarrow X_s \mu^-\mu^+]}{{\cal
B}[B\rightarrow X_s \mu^-\mu^+]_0} & =
&\left(\left[\frac{\lambda_1}{\mbox{GeV}^2}\right]\, + \,0.28 \right) \times
12.0\,\%\,.\nonumber
\end{eqnarray}
Note that the $1/m_b^2$ correction depends linearly on $\lambda_1$, therefore
these formulae are valid for all values of $\lambda_1$.
They also  imply that for $\lambda_1 =
-0.28\,\mbox{GeV}^2$ (and $\lambda_2 = 0.12\, \mbox{GeV}^2$)  one
obtains the spectator model results 
\begin{eqnarray}
\Gamma_{tot}^{-1}\, A_0^{LR} & = & 4.74 \times 10^{-6},\nonumber\\
{\cal B}[B\rightarrow X_s \mu^-\mu^+]_0 & = & 6.03  \times 10^{-6}.\nonumber
\end{eqnarray}

It is instructive to compare the size of the $1/m_b^2$ correction to
the left-right asymmetry with that of other unkown corrections or
uncertainties preventing a precise theoretical prediction of this
quantity. We looked at various sources of uncertainties,
a complete list of the corresponding errors is given in Table 1.

\begin{table}[t]\centering
\begin{tabular}{|l||r|r|}
\hline
Uncertainty due to ...& $\Delta A^{LR}/A^{LR}_0$ & $\Delta  {\cal B}/{\cal B}$ \\
\hline\hline
{\rm long-distance effects} & $+15 \%$ & +10\% \\
\hline
$m_b/2<\mu<2m_b$ & $ \pm 8 \% $ & $\pm 2 \%$ \\
\hline
$m_t = (167 \pm 6)\,{\rm GeV}$ & $\pm 6\%$ & $\pm 7 \%$ \\
\hline
$m_c = (1.4\pm 0.1)\,{\rm GeV}$ & $\pm 6\%$ & $\pm 6 \%$ \\
\hline
${\cal B}[B \rightarrow X_c l^- \overline{\nu}]$ & $\pm
4.8\%$ & $\pm 4.8\%$ \\
\hline
$\vert V_{ts}/V_{cb} \vert^2 = 0.95 \pm 0.04 $ & $\pm 4.2\%  $ & $\pm
4.2\%$  \\
\hline
$\alpha_s(M_Z)= 0.118\pm 0.003$ & $\pm 1.5\%$ & $\pm 0.2 \%$ \\
\hline
$m_b = (4.8 \pm 0.3) \, {\rm GeV}$  & $ \pm 0.7\% $ & $\pm 1.1 \%$ \\
\hline
$m_s = (0.1 \pm 0.1) \, {\rm GeV}$ & $ \pm 0.7\% $ & $\pm 0.6 \% $\\
\hline
\end{tabular}
\caption{\small Various sources of uncertainty for the integrated left-right
asymmetry as well as for the total branching ratio of the decay $B
\rightarrow X_s \mu^+ \mu^-$, in order of their numerical
importance. For the first two entries, see the explanations in the text.}
\end{table}
The most important error stems from the lack of the
next-to-next-to-leading order QCD calculation. This truncation of the
perturbative series manifests itself in the renormalization scale
dependence of the spectra and integrated rates. By varying the
renormalization scale $\mu$ between $m_b/2$ and $2 m_b$ (i.\,e.\ in
such a manner that $\ln m_b/\mu$ remains small) we estimated the
corresponding uncertainty to be about $\pm 8 \%$. Apart from this we
have to cope with our insufficient knowledge of the masses of the top
and charm quark, which enter through loop diagrams and the phase-space
factor $f(z)$. Both of these masses give rise to an uncertainty of
roughly $\pm 6 \%$. The errors due to the CKM matrix elements and the
semileptonic branching ratio are not large, but also comparable to the
$1/m_b^2$ correction. The other inaccurate known paramters, including
$\alpha_s(M_Z)$ and the masses of the remaining quarks, affect the
left-right asymmetry only by a small amount (i.\ e.\ not more than
$\pm 1 \%$). Apart from these rather trivial dependences on the
parameters, there remains one potential systematic error, namely the
long-distance contributions due to $c \bar c$-resonances. Modelling
the resonances as in Ref. \cite{kruegersehgal}, we estimated them to
enlarge the left-right asymmetry by roughly 15\% (whereby this number
depends to some extent on the applied experimental cuts). In the
worst case, lacking anything better, the associated error should be
taken to be in the same order of magnitude than the correction itself,
albeit we would call this a very conservative and perhaps unnecessary
pessimistic view.

The numbers of Table 1 show that the $1/m_b^2$ corrections are
numerically not very significant compared to all the other
uncertainties presently
existing. However, two remarks should
be made. On the one hand, we would have arrived at a much larger
correction of about 10\% if we had used the old value $\lambda_1 =
0.5\,{\rm GeV}^2$ as in Ref.\
\cite{falklukesavage94} (which was published only three years ago). It is
interesting to note that at the moment most experimental and
theoretical developments are in favour of the
assumption that the $1/m_b^2$ corrections are generically small. On
the other hand, the $1/m_b^2$ corrections are quantities of principle
interest. Assuming the validity of the HQE, there existence is a
fundamental and 
predictable property of QCD. So they should be calculated indepently of
their actual size (which, of course, is by no means known {\em a
priori}). Even if today the overall error is dominated by the
uncertainties in the SM parameters, one cannot say how this
picture will change during the forthcoming years.

\subsubsection*{Summary}
In this brief note we presented the nonperturbative
$1/m_b^2$ correction to the left-right asymmetry for the rare decay
$B\rightarrow X_s\,\mu^+\mu^-$. With a present range  of values for
the parameter $\lambda_1$ we found a correction as small as roughly
four percent for 
the integrated left-right asymmetry.
We also showed the invariant mass spectrum with these new values for
$\lambda_1$ and found the correction significantly smaller than in Ref.
\cite{falklukesavage94}.

\subsubsection*{Note added in proof}

After completion of this work a recalculation \cite{ali96} of the
$1/m_b^2$-correction to the decay $B \rightarrow X_s \, \mu^+ \mu^-$
appeared. The authors of Ref.\,\cite{ali96} use the same method
as Falk, Luke and Savage \cite{falklukesavage94} but obtain a
different result. If Ref.\,\cite{ali96} should be confirmed, our numerical
results would of course change.

\subsubsection*{Appendix: Input parameter}
$m_b=4.8\,\mbox{GeV}$,\quad
$m_c=1.4\,\mbox{GeV}$,\quad
$m_s=0.1$,\quad
$m_{\mu}=0$,\quad 
$m_t=167\,\mbox{GeV}$,\quad 
$M_W=80.2\,\mbox{GeV}$,\quad
$\mu = 5 \,\mbox{GeV}$,\quad
$\left|V_{tb}\right|=1$,\quad
$\left|V_{ts}/V_{cb}\right|=0.95$,\quad
$\sin^2\theta_W=0.23$,\quad
$\alpha =\alpha(M_Z)=1/129$,\quad
$\alpha_s(M_Z)=0.118$,\quad
${\cal B}(B\rightarrow X_c l^-\bar{\nu})=0.105$,\quad
$\lambda_2 =0.12\,\mbox{GeV}^2$,\quad
$-0.6\,\mbox{GeV}^2\leq \lambda_1 \leq -0.1\,\mbox{GeV}^2$\,(see text).


\begin{thebibliography}{10}

\bibitem{alietalfb}
{\sc A.~Ali}, {\sc T.~Mannel}, and {\sc T.~Morozumi},
\newblock {\em Phys.~Lett.} {\bf B273} (1991) 505.

\bibitem{hewett}
{\sc J.~Hewett},
\newblock {\em Phys.~Rev.} {\bf D53} (1996) 4964.

\bibitem{liudelb}
{\sc D.~Liu} and {\sc R.~Delbourgo},
\newblock {\em Phys.~Rev.} {\bf D53} (1996) 548.

\bibitem{alietalc67}
{\sc A.~Ali}, {\sc G.~F. Guidice}, and {\sc T.~Mannel},
\newblock {\em Z.~Phys.} {\bf C67} (1995) 417.

\bibitem{misiakb393}
{\sc M.~Misiak},
\newblock {\em Nucl.~Phys.} {\bf B393} (1993) 23,
\newblock ibid. \bf B439 \rm(1995) 461E.

\bibitem{burasmuenz}
{\sc A.~J. Buras} and {\sc M.~M{\"u}nz},
\newblock {\em Phys.~Rev.} {\bf D52} (1995) 186.

\bibitem{lim}
{\sc C.~S. Lim}, {\sc T.~Morozumi}, and {\sc A.~I. Sanda},
\newblock {\em Phys.~Lett.} {\bf B218} (1989) 343.

\bibitem{deshpandepanose}
{\sc N.~G. Deshpande}, {\sc J.~Trampeti\'{c}}, and {\sc K.~Panose},
\newblock {\em Phys.~Rev.} {\bf D39} (1989) 1461.

\bibitem{donnel}
{\sc P.~O'Donnel} and {\sc H.~K.~K. Tung},
\newblock {\em Phys.~Rev.} {\bf D43} (1991) R2067.

\bibitem{kruegersehgal}
{\sc F.~Kr{\"u}ger} and {\sc L.~M. Sehgal},
\newblock {\em Phys.~Lett.} {\bf B380} (1996) 199.

\bibitem{chayetal}
{\sc J.~Chay}, {\sc H.~Georgi}, and {\sc B.~Grinstein},
\newblock {\em Phys.~Lett.} {\bf B247} (1990) 399.

\bibitem{bigietal}
{\sc I.~Bigi}, {\sc N.~Ultrasev}, and {\sc A.~Vainstein},
\newblock {\em Phys.~Lett.} {\bf B293} (1992) 430,
\newblock ibid. \bf B297 \rm(1993) 477E.

\bibitem{chibisov}
{\sc B.~Chibisov}, {\sc R.~D. Dikeman}, {\sc M.~Shifman}, and {\sc
  N.~Uraltsev},
\newblock {\em preprint} TPI-MINN-96-05-T, hep-ph/9605465.

\bibitem{falklukesavage94}
{\sc A.~F. Falk}, {\sc M.~Luke}, and {\sc M.~J. Savage},
\newblock {\em Phys.~Rev.} {\bf D49} (1994) 3367.

\bibitem{neubertrev}
{\sc M.~Neubert},
\newblock {\em Phys.~Rep.} {\bf 245} (1994) 259.

\bibitem{neubert1992}
{\sc M.~Neubert},
\newblock {\em Phys.~Rev.} {\bf D46} (1992) 1076.

\bibitem{eletskii}
{\sc V.~Eletskii} and {\sc E.~Shuryak},
\newblock {\em Phys.~Lett.} {\bf B276} (1992) 191.

\bibitem{ballbraun}
{\sc P.~Ball} and {\sc V.~M. Braun},
\newblock {\em Phys.~Rev.} {\bf D49} (1994) 2472.

\bibitem{neubert9608211}
{\sc M.~Neubert},
\newblock {\em preprint {\rm CERN-TH/96-208, hep-ph/9608211}} .

\bibitem{falketal95}
{\sc A.~F. Falk}, {\sc M.~Luke}, and {\sc M.~J. Savage},
\newblock {\em Phys.~Rev.} {\bf D53} (1996) 6316.

\bibitem{gremmetal}
{\sc M.~Gremm}, {\sc A.~Kapustin}, {\sc Z.~Ligeti}, and {\sc M.~B. Wise},
\newblock {\em Phys.~Rev.~Lett.} {\bf 77} (1996) 20.

\bibitem{bigietal94}
{\sc I.~I. Bigi}, {\sc M.~A. Shifman}, {\sc N.~G. Uraltsev}, and {\sc A.~L.
  Vainshtein},
\newblock {\em Int. J. Mod. Phys.} {\bf A9} (1994) 2467.

\bibitem{bigietal95}
{\sc I.~I. Bigi}, {\sc M.~A. Shifman}, {\sc N.~G. Uraltsev}, and {\sc A.~L.
  Vainshtein},
\newblock {\em Phys.~Rev.} {\bf D52} (1995) 196.

\bibitem{kapustin}
{\sc A.~Kapustin}, {\sc Z.~Ligeti}, {\sc M.~B. Wise}, and {\sc B.~Grinstein},
\newblock {\em Phys.~Lett.} {\bf B375} (1996) 327.

\bibitem{ali96}
{\sc A.~Ali}, {\sc G.~Hiller}, {\sc L.~T. Handoko}, and {\sc T.~Morozumi},
\newblock {\em preprint} DESY 96-206, hep-ph/9609449.

\end{thebibliography}
\end{document}